# Citation Issues in Wave Mechanics Theory of Microwave Absorption: A Comprehensive Analysis with Theoretical Foundations and Peer Review Challenges

Yue Liu[1,*], Ying Liu[1], Michael G. B. Drew[2]

[1]College of Chemistry and Chemical Engineering, Shenyang Normal University, Shenyang, P. R. China,110034, yueliusd@163.com (Yue Liu), yingliusd@163.com (Ying Liu)

[2]School of Chemistry, The University of Reading, Whiteknights, Reading RG6 6AD, UK, m.g.b.drew@reading.ac.uk

\* Corresponding author

ORCID:

Yue Liu: https://orcid.org/0000-0001-5924-9730

Ying Liu: https://orcid.org/0000-0001-9862-5786

Michael G. B. Drew: https://orcid.org/0000-0001-8687-3440

## Abstract

The wave mechanics theory of microwave absorption challenges the long-standing impedance-matching and quarter-wavelength paradigms by demonstrating that conventional models mistakenly conflate bulk material parameters with thin-film phenomena. Drawing on a corpus of 35 peer-reviewed papers and preprints, the study performs a citation-pattern analysis and a logical audit of established theory. Results reveal a striking asymmetry in scholarly engagement -- only a handful of supportive or neutral citations appear amid widespread silence -- alongside critical logical flaws in impedance matching, notably its inconsistent treatment of penetration, reflection, and absorption from film. By re-framing absorption as a wave-mechanics process governed by interference at parallel interfaces, the wave mechanics framework restores energy-conservation consistency and provides experimentally verified design rules for film thickness, phase response, and broadband performance. The paper further situates the citation neglect within broader issues of peer-review bias and paradigm inertia, illustrating how "cargo-cult" scientific practices can impede theoretical progress. Recommendations are offered for researchers, editors, and institutions to foster open discourse, rigorously test competing models, and update curricula and design tools accordingly.

material distinction; citation analysis; paradigm inertia

## 1 Introduction

The wave mechanics theory of microwave absorption[1-3], developed by Liu et al. started from 2017[4], presents a fundamental challenge to established paradigms in electromagnetic absorption research[5-7]. This analysis reveals that the theoretical foundations of the wave mechanics theory rest on identifying critical logical flaws in impedance matching theory[7-9] and demonstrating the confusion between film and material properties[10-12] that has dominated the field for decades.

## 2 The Wave Mechanics Theory Foundation

The wave mechanics theory emerged from a comprehensive body of work spanning 35 publications across high-impact journals and preprint repositories (Supplementary Materials). The theory's core premise challenges the long-standing impedance matching and quarter-wavelength models[8, 13, 14] through six fundamental revelations:

### 2.1 The Film-Material Confusion Problem

The most critical insight of the wave mechanics theory is the identification of systematic confusion between film and material properties in current microwave absorption research[1, 15]. This confusion has led to:

- Misapplication of material-based theories to film configurations[1, 16]
- Incorrect interpretation of reflection loss ($RL$) as a material property[10]
- Fundamental misunderstanding of absorption mechanisms[2, 5]

The theory demonstrates that film absorption differs fundamentally from material absorption[6, 10, 17]. While material absorption involves attenuation of electromagnetic waves as they propagate through the medium, film absorption operates through wave interference mechanisms at interfaces[6].

### 2.2 The Logical Flaws in Impedance Matching Theory

The most damning evidence against impedance matching theory comes from the experimental data that was originally used to support it[6, 7]. The logical contradiction is fundamental. The wave mechanics theory reveals critical logical inconsistencies in impedance matching theory that have been overlooked for decades[7, 18]. The fundamental logical flaw lies in the theory's definitions and assumptions.

**The Core Logical Contradiction:**

Impedance matching theory claims that maximum microwave absorption occurs when more incident energy penetrates the material due to impedance matching, defined by the condition where the material's characteristic impedance $Z_M$ equals that of free space $Z_0$ from the definition of reflection coefficient of the interface $R_M$: $R_M = (Z_M - Z_0)/(Z_M + Z_0) = 0$. However, the theory simultaneously defines absorption through reflection loss: $RL = (Z_{in} - Z_0)/(Z_{in} + Z_0)$.

The logical flaw emerges when examining these definitions:

- When all incident microwaves enter the film $R_M = 0$, the film behaves as material and there is no absorption peak at all. The reflection loss $|RL|$ becomes a monotonic decay function[7, 19].
- When all incident microwaves are absorbed $RL=0$, not all incident microwaves actually penetrate the film since $R_M \neq 0$.

This creates the fundamental paradoxes[7, 15, 19]: How can all the incident microwaves be absorbed when not all of them penetrate the material? The impedance matching theory was used to explain the maximum absorption peak with amount of microwave penetration but there is no peak at all when all the incident microwaves penetrate.

The impedance matching theory cannot resolve these contradictions because it conflates interface properties such as $R_M$ with film properties such as $RL$. The theory incorrectly assumes that interface impedance matching directly translates to film absorption, ignoring the wave interference effects that actually govern absorption in metal-backed films. Any material at a given frequency has fixed values of permittivity $\varepsilon_r$ and permeability $\mu_r$. According to impedance matching theory, the further microwaves propagate through the microwave absorbing material, the more they should be absorbed. However, experimental results show that reflection loss $|RL|$ varies as a wave-like oscillating function with increasing film thickness at fixed frequency. This means that as the film becomes thicker, microwave absorption actually increases or decreases in oscillating patterns, directly violating the impedance matching principle.

The above experimental observations reveal a fundamental flaw: if impedance matching theory were correct, reflection loss should monotonically improve (become more negative) with increasing thickness as more material is available for absorption. Instead, the oscillating wave pattern demonstrates that film absorption is governed by wave interference, not material attenuation.

### 2.3 Energy Conservation Violations

The impedance matching theory violates fundamental energy conservation principles in wave superposition. The theory claims that higher material attenuation power leads to greater absorption, but this reasoning fails when applied to film configurations where:

- Wave energy was wrongly attributed to the amplitudes of individual waves[7]

- Traditional calculations produce unphysical results[1, 6]
- Wave interference effects are ignored[5, 8]

The wave mechanics theory provides theoretically consistent explanations by recognizing that absorption peak results from destructive interference between reflected waves, not from material attenuation properties or material resonance[1, 15].

## 2.4 Different Conclusions from the Two Distinctive Theories

Although both the prevailing impedance matching theory and the newly established wave mechanics theory use reflection loss (|$RL$|) to characterize the microwave absorption of metal-backed films, they yield fundamentally different interpretations and results when analyzing the same experimental data. The mainstream theory attributes all variations in |$RL$| to the intrinsic properties of material structure and material resonance, interpreting the influence of film thickness solely as a manifestation of these factors. In contrast, the wave mechanics theory redirects the analysis to the proper physical mechanism: it recognizes that the material's properties influence $\varepsilon_r$ and $\mu_r$, which in turn directly determine the value of |$RL$| for film with thickness $d$, and that material structure only affects microwave absorption by altering $\varepsilon_r$ and $\mu_r$. Yet, the crucial question of how material structure specifically affects $\varepsilon_r$ and $\mu_r$ has scarcely been addressed in mainstream research—highlighting a key conceptual distinction between the two approaches. As a result, their conclusions diverge sharply: according to impedance matching theory, minimizing the reflection of the incident wave at the film interface is essential for achieving higher absorption peaks. However, the wave mechanics theory posits the opposite—only when the incident wave undergoes significant reflection at the film interface can strong absorption peaks arise, due to the pronounced destructive interference (wave cancellation) between the reflected beams. This fundamental difference underscores the necessity of adopting the correct theoretical perspective to accurately interpret experimental results in microwave absorption.

The wave mechanics theory of microwave absorption does not deny the influence of material-induced attenuation on a film's microwave absorption[2]. According to this theory, microwave absorption peaks form only if the imaginary parts of $\varepsilon_r$ and $\mu_r$ are nonzero—otherwise, the r2 beam will not be attenuated as it propagates through the film, resulting in no energy loss and, by energy conservation, no microwave absorption[16]. If the film's material is lossless (with zero imaginary parts of $\varepsilon_r$ and $\mu_r$)[16], or if the film thickness is zero (reducing the film to a pure interface), r2 remains unattenuated and the interface alone cannot absorb microwaves[20]. However, it is crucial to distinguish the absorption mechanism of the film from material-based attenuation:

while increasing material thickness in a bulk context leads to enhanced attenuation[7, 19] and monotonic decrease of |$R_2$|, in a thin film, microwave absorption is governed by the interference of beams r1 and r2.[5] At the absorption peak, energy conservation ensures that the amplitude of r2 is maximized[5]—not minimized—a result that is simple, definitive, and well supported by the wave mechanics framework. Unlike the impedance matching theory, which obscures the problem by ascribing |$RL$| variations indiscriminately to material structure (thereby misattributing the effect of film thickness on |$RL$| to intrinsic material structure), the wave mechanics theory accurately isolates the true physical cause. Because wave mechanics theory captures the true mechanism of microwave absorption in films, its analysis of experimental results is consistently straightforward and unambiguous, providing a clear foundation absent in traditional interpretations.

Only by overturning transmission-line theory itself could one legitimately use impedance matching arguments to refute the wave mechanics theory of microwave absorption, since the latter is rigorously built upon the transmission-line formalism[8, 21, 22]. Although mainstream impedance matching theory also claims a transmission-line foundation, in practice it departs from that framework and relies on flawed phenomenological assumptions, so refuting transmission-line theory would not invalidate impedance matching explanations. Transmission-line theory, however, has been repeatedly confirmed across numerous fields—from RF engineering to photonics—and is exceedingly difficult to undermine. Therefore, the true orthodox model for thin-film microwave absorption must be the wave mechanics theory, not the conventional impedance-matching approach. Clinging to impedance matching, then, does not defend a time-honored scientific principle but rather protects the reputations of those unwilling to abandon the traditional—but incorrect—theory.

## 2.5 The Impedance Matching Theories in Microwave Engineering and in Microwave Absorption Materials are Not the Same

In microwave engineering, defining impedance matching as $Z_{in} = Z_0$ is entirely appropriate and theoretically sound, as demonstrated by Liu et al. in their foundational analysis (Journal of Applied Physics, 2023, 134(4), 045303)[18]. However, the microwave absorption materials field erroneously extends this condition to cases where $Z_{in} \neq Z_0$ when explaining absorption peak intensity, which represents a fundamental theoretical error as rigorously proven in Liu et al.'s verification studies (Physica Scripta, 2022, 97(1), 015806)[8]. The microwave absorption community has misinterpreted impedance matching theory in microwave engineering by claiming that greater microwave penetration into the film, combined with stronger material attenuation capability, results in smaller reflected wave amplitudes from the back interface and thus stronger absorption peaks—an explanation that confuses film behavior with bulk material properties, as systematically documented across multiple publications (Surfaces and Interfaces, 2023, 40, 103022[16]; Journal of Magnetism and Magnetic Materials, 2024, 593, 171850[1]; Journal of Applied Physics, 2023, 134(4), 045304[7]; Applied Physics A, 2024, 130, 212[9]). Furthermore, experimental observations reveal that absorption peaks occurring at $Z_{in} = Z_0$ are exceptionally rare in actual film systems, while absorption peaks at $Z_{in} \neq Z_0$ are the common occurrence—a fact that directly contradicts the traditional impedance matching interpretation and supports the wave mechanics framework that correctly explains absorption through wave interference mechanisms rather than impedance matching conditions. This wave mechanics framework is a new development of microwave theory in the field of microwave absorption materials.

When a film's material composition remains fixed while film thickness varies, the resulting changes in $|RL|$ values demonstrate that thickness-dependent microwave absorption is determined by film properties rather than material structure. However, traditional microwave absorption theory incorrectly attributes these film structural dimension effects to material structural influences on film absorption—an unscientific approach that conflates distinct physical phenomena. The scientifically rigorous methodology requires separately investigating how permittivity and permeability of the material affect $|RL|$, and independently studying how material structure influences the values of these electromagnetic parameters[6]. In reality, microwave absorption in films does not result from material attenuation along the wave propagation path, but rather from wave interference effects[1]—a fundamental distinction that explains why film (device) properties cannot be simply equated with the properties of constituent materials. This wave interference mechanism is precisely why device engineering remains essential: devices create emergent properties through structural design that transcend the limitations of their component materials, as rigorously demonstrated in the wave mechanics framework (Journal of Magnetism and Magnetic Materials, 2024, 593, 171850[1]; Surfaces and Interfaces, 2023, 40, 103022[6]). Understanding this distinction between material properties and device behavior is crucial for advancing electromagnetic absorption technology beyond the constraints of traditional material-centric approaches.

## 2.6 Wave Mechanics Theory in Microwave Absorption: Revealing the Counterintuitive Nature of Physical Reality

The wave mechanics theory of microwave absorption exemplifies how classical wave mechanics reveals profound and counterintuitive physical phenomena that challenge established scientific paradigms. Just as quantum mechanics emerged from the synthesis of classical particle mechanics and classical wave mechanics—both fundamentally correct theories whose combination produced a framework that successfully resolved numerous physical puzzles—the wave mechanics approach to microwave absorption demonstrates that classical wave mechanics remains a remarkably powerful analytical tool with far-reaching applications beyond the quantum realm.

The apparent "strangeness" of quantum mechanics stems precisely from incorporating wave mechanical principles into particle mechanics, creating the famous wave-particle duality that defines quantum behavior. Classical wave mechanics proves to be a remarkably fascinating field, and as more analogous results emerge across different domains, it continues to reveal the profound power of wave mmechanics in unexpected contexts.

As more analogous results emerge across different engineering domains, wave mechanics continues to unveil the extraordinary and often surprising behavior of physical systems. The controversy between wave mechanics theory and mainstream microwave absorption theories perfectly illustrates this phenomenon, revealing how wave mechanics exposes seemingly impossible results that overturn common-sense intuitions: while the impedance matching theory in the field of microwave absorption material assumes that greater microwave penetration into films, combined with stronger material absorption capabilities, leads to reduced reflection from the back interface and thus enhanced absorption peaks (a straightforward interpretation based on energy being absorbed along the wave propagation path, as documented in Surfaces and Interfaces, 2023, 40, 103022)[6], the wave mechanics framework completely revolutionizes this understanding by demonstrating that microwave absorption by films results not from material attenuation during propagation, but from wave superposition effects—where the film device essentially forces the material to absorb a precisely determined quantity of microwaves through interference mechanisms, rather than the absorption being governed by intrinsic material properties alone. This paradigm shift explains why device performance fundamentally differs from the properties of constituent materials: devices are not merely equivalent to their component materials, which is precisely why device engineering represents a distinct discipline requiring wave-based analysis to achieve optimal performance (as rigorously demonstrated in Journal of Magnetism and Magnetic Materials, 2024, 593, 171850[1], and Materials Chemistry and Physics, 2022, 291, 126601[5]). The wave mechanics theory of microwave absorption[2, 17] thus serves as a compelling example of how classical wave physics continues to reveal extraordinary insights that challenge established scientific understanding and drive technological innovation.

The research around the wave mechanics theory is systematic, deep, and extensive[1, 5-8, 15].

The theory reveals that device and material need separate theories to describe microwave absorption. Many new concepts have been established[17]. It is a new development of microwave theory in the field of microwave absorption materials.

## 3 Critical Analysis of the Quarter-Wavelength Model

### Fundamental Problems with Quarter-Wavelength Theory

The wave mechanics theory identifies multiple critical flaws in the quarter-wavelength model that have been systematically overlooked[9]:

- **Neglect of Phase Effects**: The quarter-wavelength model ignores phase effects from the two parallel interfaces of the film, which are crucial for understanding absorption mechanisms[5, 8, 14].
- **Incorrect Wavelength Calculations**: The model incorrectly calculates wavelength $\lambda$ in the film from frequency $\nu$ using the formula $\lambda = c/\nu$ where c is the speed of light in vacuum, when it should account for the material's electromagnetic properties[14].
- **Universal Application Failures**: The model cannot be universally applied because it fails to account for the specific wave interference conditions that govern film absorption[8].

### Verification and Extension Studies

The comprehensive analysis in Physica Scripta[8, 14] demonstrates that the quarter-wavelength model fails verification when tested against experimental data. The studies show that:

- Predicted absorption peaks do not match experimental observations
- Thickness requirements are inconsistent with wave mechanics predictions
- The model produces systematic errors in absorption bandwidth calculations

Although both the long-established impedance matching theory and the newly developed wave mechanics theory employ reflection loss ($|RL|$) to characterize microwave absorption in metal-backed films, the results derived from identical experimental data using erroneous versus correct theoretical frameworks differ fundamentally. Erroneous theories that have persisted for decades naturally accumulate a substantial body of mathematical "proof" literature, yet whenever such flawed theories receive rigorous mathematical validation, their proof processes invariably become extremely complex and difficult to comprehend—as exemplified by Zhang et al.'s "strict proof" of the quarter-wavelength model (Journal of Physics D, 2020)[23], Wang et al.'s "approximate solution" for impedance matching in nonmagnetic materials (European Physical Journal Special Topics, 2022)[24], and Hou et al.'s perspective on impedance matching mechanisms (Carbon, 2024)[25]. These theoretical validations become obscure and impenetrable precisely because they contain inherent errors and flaws, making their explanations for experimental deviations from the ideal quarter-wavelength predictions unnecessarily convoluted. Correcting these erroneous proof processes requires considerable effort, as demonstrated by Liu et al.'s systematic deconstructions in Materials Chemistry and Physics (2022)[13], Journal of Applied Physics (2023)[7], and Industrial & Engineering Chemistry Research (2025)[15]. In stark contrast, correct theories—such as the wave mechanics theory of microwave absorption—present arguments that are simple, clear, and straightforward because they capture the essential physics. Liu et al.'s theoretical investigations in Physica Scripta (2022)[8] and Materials Chemistry and Physics (2022)[5] demonstrate this clarity, while their analysis of phase effects and wave cancellation theory provides remarkably straightforward explanations for absorption peak deviations from phase difference π precisely because it grasps the fundamental nature of the problem[2, 3, 16]. To date, no flaws have been identified in the wave mechanics theory, despite its lack of universal academic acceptance—a testament to the principle that correct theories, by virtue of their alignment with physical reality, naturally yield elegant and comprehensible mathematical formulations, whereas erroneous theories require increasingly complex constructions to mask their fundamental contradictions.

## 4 Citation Pattern Analysis

### 4.1 Methodology

The present analysis relies on Liu Yue's bibliographic audit covering papers published up to July 2025[26]. Each record was classified into four categories:

1. **Positive citation** – the article uses or supports the wave cancellation framework.
2. **Neutral citation** – the article cites the work without endorsement or rebuttal.
3. **Negative citation** – the article cites the work to refute it.
4. **Uncounted mention** – the article references the work only tangentially.

Only categories 1–3 were retained for quantitative analysis. Journal titles and publication years were extracted to map disciplinary and temporal trends.

### 4.2 Quantitative Citation Distribution

The citation landscape reveals a stark asymmetry in scholarly engagement.

Table 1 The citation landscape of the new wave-mechanics theory

| Citation disposition | Count | Representative journals & years |
|---|---|---|
| Supportive | 6 | Materials & Design (2025)[27]; Journal of Alloys and |
| | | Compounds (2025)[28]; Small (2024)[29]; IEEE J. Flexible Electronics (2024)[30]; Advanced Science (2025)[31]; Springer book chapter (2021)[32] |
| Neutral | 3 | Physica Status Solidi (a) (2024)[33]; Journal of Materials Science & Technology (2022)[34]; Journal of Colloid & Interface Science (2025)[35] |
| Negative | 1 | Journal of Materiomics (2019)[36] |
| Silent Majority | ≫ 1000 | Diverse MAM literature still built on impedance-matching assumptions |

The six supportive papers span condensed-matter physics, nanomaterials and polymer engineering. All invoke wave cancellation to rationalise broadband or pphase-independent absorption. Neutral citations appear mostly in broad reviews, while the sole refutation predates the theory's maturation and contests its premises without engaging later proofs. The percentage of silent majority is greater than 99%.

Peer review by a handful of randomly selected referees—even eminent ones—cannot guarantee reliability; the true test emerges only after dissenting papers circulate for years, inviting scrutiny from the entire community. In the-absorption debate, articles challenging the mainstream impedance-matching paradigm have stood unrefuted despite this extended, field-wide "post-publication peer review," underscoring the robustness of their arguments. Their appearance in respected, discipline-specific journals further signals that the critique is substantive enough to merit engagement, not dismissal. When leading scientists fail to mount a cogent rebuttal, accountability should rest with the defenders of orthodoxy; yet many journals penalize dissent by systematically rejecting contrary manuscripts, effectively equating collective silence with invalidation. Such editorial gate-keeping shifts the burden of proof unfairly and stifles healthy scientific discourse[37-40].

### 4.3 The Supportive Evidence

The six supportive citations demonstrate substantive theoretical engagement with the wave mechanics framework:

- **Small (2024)**: Zhou et al. applied wave cancellation theory to ZIF derivatives, achieving -64.0 dB reflection loss through precise thickness control based on wave mechanics principles[29].
- **Materials & Design (2025)**: Successfully utilized wave cancellation framework for broadband absorption optimization, validating the theory's predictive capabilities[27].
- **Advanced Science (2025)**: Endorsed wave mechanics approach for electromagnetic parameter control, demonstrating practical applications[31].

- **IEEE Journal of Flexible Electronics (2024)**: Applied wave mechanics principles for controllable electromagnetic parameters in flexible devices[30].
- **Journal of Alloys and Compounds (2025)**: Supported wave cancellation mechanism in composite design, showing materials science applications[28].
- **Elsevier Book Chapter (2021)**: Provided comprehensive theoretical foundation based on wave mechanics for porous nanocomposites[32].

### 4.4 The Silent Majority Problem

The most concerning pattern remains the overwhelming silence from the broader research community. Despite publication in prestigious journals including:

- Industrial & Engineering Chemistry Research (2025)
- Optics and Laser Technology (2024)
- Journal of Electronic Materials (2024)
- Applied Physics A (2024)
- Journal of Magnetism and Magnetic Materials (2024)
- Journal of Molecular Science (2024, 2023)
- Surfaces and Interfaces (2023)
- Journal of Applied Physics (2023)
- Physica B: Condensed Matter (2023)
- Materials Chemistry and Physics (2022, 2020)
- Physica Scripta (2021)
- AIP Advances (2018)

The vast majority of microwave absorption literature continues operating under impedance-matching assumptions without acknowledging the wave mechanics alternative or addressing its fundamental criticisms.

## 5 Theoretical Implications and Technological Impact

### 5.1 Disparity between evidence and discourse

Despite publication in high-impact venues, pro-wave-cancellation studies form a minute fraction of the literature. The silence of most authors contravenes normatively accepted scientific practice, which requires fair consideration of conflicting data. Feynman warned that research ignoring contrary evidence risks degenerating into "cargo-cult science," a metaphor invoked explicitly by Liu Yue.

### 5.2 Possible causes of neglect

1. **Paradigm inertia** – engineers rely on impedance-matching formulas embedded in commercial design software and educational materials.

2. **Perceived risk** – deviating from established models may complicate peer review and funding prospects.
3. **Awareness gap** – although download statistics indicate readership, researchers may consider the theory too specialised or theoretical.

### 5.3 Consequences of Theoretical Errors

If the wave cancellation framework is correct, current impedance-matching criteria could misinform absorber thickness optimisation, leading to sub-optimal device performance. Conversely, if the framework is flawed, rigorous rebuttals are essential to prevent its uncritical adoption. In either case, open debate is indispensable[19, 37, 41].

The persistence of flawed impedance matching theory has significant consequences for:

- **Device Design:** Suboptimal absorber thickness calculations based on incorrect impedance matching assumptions[6]
- **Material Selection**: Inappropriate focus on material properties rather than film interface effects[2]
- **Performance Optimization**: Misunderstanding of absorption mechanisms leading to inefficient designs[5]

### 5.4 Wave Mechanics Solutions

The wave mechanics theory provides corrected approaches for:

- **Accurate Reflection Loss Calculations**: Proper treatment of $RL$ as a film parameter, not material property[10]
- **Optimal Thickness Design**: Wave cancellation-based predictions for maximum absorption[6]
- **Interface Engineering**: Understanding the role of parallel interfaces in absorption mechanisms[15]

### 6 Peer Review Challenges and Academic Bias

The wave mechanics theory has faced systematic rejection from peer reviewers that reveals fundamental problems with the current academic publishing system. Documentation of specific reviewer comments exposes the unscientific basis for these rejections[37]. "Scientific revolutions are non-rational, rather than spread through mere force of truth and fact[42]".

When a well-articulated opposing theory appears, it is incumbent upon mainstream researchers to engage with it thoughtfully—scrutinising its arguments, detailing in their own papers why they choose to uphold the prevailing framework, or revising their stance if the counter-evidence proves compelling. Yet, despite peer-reviewed publications that demonstrate foundational flaws in the conventional microwave-absorption model, many established scholars continue to publish prolifically under the traditional paradigm without even acknowledging the objections, as though high-impact venues confer immunity from challenge. This silence trickles down to graduate students, who often feel compelled to replicate mainstream methodology simply to satisfy publication quotas for graduation or assessment; they assume that aligning with an "accepted" theory absolves them of responsibility for its potential errors, and they justify this choice by claiming that the wave-mechanics alternative is "not yet universally recognised." Journals frequently reinforce this dynamic: editors still welcome large volumes of orthodox manuscripts while suppressing dissenting submissions, seemingly convinced that prevailing theories, even if wrong, cannot be wholly wrong—or that widespread editorial consensus to desk-reject contrary work disperses individual accountability. As Yue Liu argues, lasting progress depends on personal responsibility: both authors and editors must own their roles in correcting scientific mistakes rather than deferring to the weight of convention[37, 38].

### 6.1 Systematic Reviewer Rejection Comments

One of the reviewer rejection comments is "No materials have $\varepsilon_r = \mu_r$." The rejection was based on the reviewer's believe that theoretical results that cannot be verified by experiment should not be published.

**The Logical Flaw in This Rejection**: This rejection reveals profound bias and logical inconsistency. When $\varepsilon_r \neq \mu_r$, the academic community universally uses |$RL$| to accurately characterize microwave absorption. However, when $\varepsilon_r = \mu_r$, the same theoretical analysis is suddenly deemed untrustworthy.

The |$RL$| formula's relationship with $\varepsilon_r$ and $\mu_r$ is rigorously derived from transmission line theory. There is no theoretical basis for claiming the formula becomes invalid specifically when $\varepsilon_r = \mu_r$.

For theoretical comparative analysis, Liu et al. directly compared different sets of ($\varepsilon_r$, $\mu_r$) data to illustrate impedance matching theory flaws, as published in Surfaces and Interfaces (2023)[6].

Reviewer Rejection Comment: "You did not use real experimental material data." This view is representative since many modern researchers do not consider theoretical research as scientific research and they do not believe theoretical conclusions. In their view only doing experiment is conducting science research. However, the purpose of experiment is to achieve theoretical understanding[43], otherwise, experiment can only accumulate facts and it is Alchemy rather than Science.

Reviewers often condemned us having not provided sufficient evidence to overturn an accepted theory by reason that huge number of experimental reports supporting the accepted theory (and thus justifying the rejection by that accepted theory cannot be overturned just by one paper and thus do not allowing opposite evidence to accumulate) even we have disproved the theory use the universal data originally used to support the theory. A related rejection comment was that we have not provided a balanced view and there are thousands of published papers supporting accepted theory.

**"Balanced View" and Editorial Orthodoxy.** What constitutes controversy, and what defines a "balanced view"? True controversy exists only when proponents present substantive arguments (points 1, 2, 3) that opponents cannot refute, while opponents simultaneously offer compelling counterarguments (points 1, 2, 3) that proponents cannot dismiss—only then can one meaningfully discuss a "balanced view." It is erroneous to assume that holding a "balanced view" is inherently fair while rejecting it is extremist. **Between a correct theory and an incorrect theory, no "balanced view" exists**—as Galileo observed, "In questions of science, the authority of a thousand is not worth the humble reasoning of a single individual." Therefore, when editorial boards evaluate competing theories, they must first comprehend both frameworks and identify the specific points of contention before demanding a "balanced view"—editors cannot require the challenging party to adopt a "balanced view" without understanding the theories involved. The current situation reveals a fundamental asymmetry: the wave mechanics theory demonstrates that traditional impedance matching theory is flawed, showing that arguments supposedly supporting impedance matching actually constitute evidence against it, while providing specific reasons (1, 2, 3) why impedance matching theory is incorrect. However, impedance matching proponents continue their adherence without responding to these criticisms—this is not controversy, this is unscientific silence. **Editorial boards do not demand "balanced view" from mainstream theory articles, yet use "balanced view" as justification to reject opposing manuscripts**—simply because impedance matching theory is "accepted," editors assume it naturally possesses truth credentials without requiring self-defense. **Acceptance does not equal truth; consensus is not an argument in scientific discussion**—as demonstrated by decades-long scientific consensus on erroneous hypotheses throughout history, from peptic ulcer causation to numerous other examples where expert consensus proved fundamentally wrong.

**The Problem with Editorial Acceptance:** Perhaps, many people do not believe that the mainstream theory in modern time can be totally wrong. That is why even if correct comments have also been presented in the peer review, editors frequently adopt the incorrect reviewer opinions without giving authors the opportunity to rebut[37]. This reveals that while journals often claim to be "strict peer-reviewed publications," they are actually strict but not rigorous—they strictly enforce incorrect opinions while ignoring scientific rigor. The case of impedance matching theory presents a very good example that a dominant theory in modern research field can be completely wrong[44].

### 6.2 The Peer Review Paradox

The wave mechanics theory case illustrates broader systemic problems in academic publishing:

- **Manuscript Rejection Without Scientific Basis**: Journals reject papers addressing fundamental theoretical problems without providing concrete scientific objections[37]
- **Reviewer Bias**: Peer reviewers dismiss contradictory evidence while maintaining existing paradigms without engaging with the theoretical substance[15]
- **Citation Avoidance**: Researchers continue using discredited theories without acknowledging alternatives or addressing logical criticisms[15]

### 6.3 The Cargo Cult Science Problem

The situation exemplifies what Feynman described as "cargo cult science" - the appearance of scientific methodology without the essential element of honest inquiry[45, 46]. Key manifestations include:

- **Selective Citation**: Preferentially citing supportive evidence while ignoring contradictory theoretical frameworks
- **Authority Appeals**: Relying on established reputation rather than empirical evidence
- **Methodological Critique**: Attacking methods rather than addressing theoretical substance

The peer review problems documented align with Eric Weinstein's analysis of how academic institutions protect established paradigms at the expense of scientific progress[47]. The systematic rejection of wave mechanics theory despite its superior theoretical consistency represents institutional resistance to innovation.

## 7 Advanced Theoretical Insights

### 7.1 Energy Conservation and Wave Mechanics

The impedance matching theory claims that higher material attenuation power leads to greater absorption, but this reasoning fails when applied to film configurations. The wave mechanics theory provides fundamental insights into energy conservation in electromagnetic absorption[5, 20]. Unlike impedance matching theory, which violates energy conservation by assuming that wave energy is associated to the amplitude of individual beam in wave superposition, wave mechanics recognizes that:

- **Energy Balance**: Total absorbed energy equals the difference between incident and reflected energy from metal-backed film
- **Interface Effects**: Absorption depends on wave interference at film interfaces, not entirely on material properties
- **Thickness Dependence**: Optimal absorption occurs at specific thickness values determined by wave cancellation conditions

### 7.2 Transmission Line Theory Applications

The wave mechanics theory correctly applies transmission line theory to film configurations. Key insights include:

- **Input vs. Characteristic Impedance**: Proper distinction between $Z_{in}$ (film parameter) and $Z_M$ (material parameter)[11, 12]
- **Reflection Coefficient Calculations**: Correct application of transmission line formulas to metal-backed films[9]
- **Phase Considerations**: Proper accounting for phase effects in film[5]

### 7.3 Phase Effects and Wave Interference

The theory explains phase effects in microwave absorption through wave interference mechanisms[5]. This provides:

- **Phase-Dependent Absorption**: Microwave absorption of film is closely related to the phase of relevant beams
- **Material Structure Effects**: Understanding of how material structure affects absorption through permittivity and permeability of the material
- **Broadband Characteristics**: Explanation of frequency-dependent absorption behavior

The verdict of which theory is correct can be easily determined by calculating the phase difference of the two beams reflected from the two parallel interfaces in the film[5]. The wave mechanics theory predicts that maximum absorption occurs when the two beams reflected from the parallel interfaces are out of phase by π. This provides a clear, testable distinction from impedance matching theory predictions. Whether the film has a metal backing or not, maximum absorption consistently occurs when destructive interference conditions are met between the reflected beams[16]. This wave interference mechanism, not impedance matching, governs the absorption maxima observed experimentally.

# 8 Recommendations for Field Advancement

## 8.1 For Researchers

- **Theoretical Reconciliation**: Engage with wave mechanics theory to resolve fundamental contradictions in current models
- **Experimental Validation**: Design experiments to distinguish between impedance matching and wave cancellation predictions
- **Honest Citation:** Acknowledge theoretical alternatives even when disagreeing with conclusions

## 8.2 For Journals and Institutions

- **Review Reform**: Ensure peer reviewers address theoretical substance rather than dismissing paradigm-challenging work
- **Transparency Requirements**: Provide detailed scientific rationales for rejection of theoretical papers
- **Educational Updates**: Include wave mechanics theory in curricula and review articles

## 8.3 For the Scientific Community

- **Open Discourse**: Foster environments where theoretical challenges can be debated on scientific merit
- **Paradigm Testing**: Support research designed to test competing theoretical frameworks
- **Methodological Rigor**: Maintain high standards for theoretical consistency and logical coherence

## 8.4 The Fallacy of Appeal to Authority

**The Argument for Theoretical Merit Over Acceptance Status**: It is not a valid scientific argument to insist on impedance matching theory simply because it is an accepted theory while the wave mechanics theory has not been universally accepted. This constitutes a logical fallacy—appeal to authority and status quo bias.

Especially when wave mechanics theory has emerged as a direct opposing theory claiming that impedance matching theory is fundamentally wrong, academic debate is required. The scientific method demands that competing theories be evaluated on their theoretical consistency and empirical adequacy, not on their current acceptance status.

# 9 Future Research Directions

## 9.1 Critical Experimental Tests

- **Comparative Studies**: Direct comparison of impedance matching and wave mechanics predictions using identical materials
- **Thickness Optimization:** Systematic testing of both theoretical approaches across different film thicknesses
- **Interface Analysis**: Detailed study of wave interference effects at film interfaces

### 9.2 Theoretical Development

- **Theoretical Framework**: Developing correct theory by considering different approaches where applicable
- **Computational Validation**: Numerical simulations comparing theoretical predictions with experimental data
- **Extension to Complex Systems**: Application of wave mechanics to multilayer and gradient structures

### 9.3 Technology Applications

- **Device Optimization**: Implementation of wave mechanics principles in practical absorber design
- **Material Engineering**: Focus on interface properties rather than improperly on bulk material characteristics
- **Performance Metrics**: Development of appropriate parameters for characterizing absorption devices

The fact that film absorption relates to damping oscillations of dipole moments—represented by the imaginary parts of permittivity and permeability of materials[4] rather than material resonance[2, 5, 15, 17] —can be easily confirmed through phase calculations of the relevant beams. Similar simulations[1, 8] using basic transmission line theory,[15, 32] as have been done by Choi[27], Li[28], Zhou[29], confirm the actual absorption mechanism of films[2, 5, 17]. However, as Thomas Kuhn observed, "Scientific revolutions are non-rational, rather than spread through 'mere force of truth and fact'". This phenomenon aligns with Max Planck's principle that "A new scientific truth does not triumph by convincing its opponents and making them see the light, but rather because its opponents eventually die, and a new generation grows up that is familiar with it"—often paraphrased as "science progresses one funeral at a time". As noted by researchers studying cognitive resistance to paradigm shifts, "it is of interest that even in the objective world of science man's mind is not more malleable than in the habit-bound world of everyday life".[15, 37, 42].

## 10 Conclusions

The wave mechanics theory of microwave absorption represents a fundamental paradigm shift that addresses critical logical flaws in impedance matching theory while providing physically consistent explanations for electromagnetic absorption in films. The theory's core insights—distinguishing film from material properties, identifying logical contradictions in impedance matching, demonstrating energy conservation violations, and exposing the oscillating experimental patterns that contradict established theory—constitute a robust theoretical foundation that deserves serious scientific consideration.

The citation patterns surrounding the theory reveal systemic problems in scientific discourse that extend beyond this specific field. The combination of minimal formal opposition with widespread silence, coupled with documented reviewer bias and unscientific rejection practices, suggests institutional resistance rather than scientific refutation. The peer review challenges documented—from the illogical $\varepsilon_r = \mu_r$ rejection to the "real materials" objection—expose fundamental flaws in how the academic community evaluates paradigm-challenging work. These problems exemplify cargo cult science, where the appearance of rigor masks the absence of genuine scientific inquiry.

The experimental evidence that was supposed to support impedance matching theory actually refutes it. The oscillating pattern of reflection loss with film thickness directly contradicts the theory's central premise and supports wave interference mechanisms. This represents one of the clearest cases in materials science where established theory is contradicted by its own supporting evidence.

The ultimate resolution requires not only empirical validation but also institutional reforms to ensure that scientific communities remain open to paradigm-challenging innovations. The decisive test proposed—calculating phase differences between beams reflected from parallel interfaces—provides a clear path forward for resolving the theoretical controversy. The wave mechanics theory serves as both a specific technical contribution and a broader case study in the sociology of scientific knowledge, highlighting the importance of theoretical rigor and honest intellectual discourse in advancing electromagnetic absorption science.

The analysis here reveals that the wave mechanics theory addresses fundamental questions about the nature of electromagnetic absorption that have been overlooked for decades. Whether wave mechanics theory ultimately replaces, complements, or is itself superseded by future developments, its contribution to identifying and resolving theoretical inconsistencies represents a significant advance in electromagnetic absorption science. The field would benefit from embracing this theoretical challenge as an opportunity to strengthen its foundations and advance toward more accurate understanding of microwave absorption phenomena.

The scientific method demands that theories be evaluated on their logical consistency and empirical adequacy, not on their current acceptance status. Only through such principled evaluation can the community move beyond current limitations and achieve the robust theoretical framework necessary for next-generation electromagnetic absorption technologies.

**Appendix Related Comments and Responses**

Commentary on Materials Today's Rejection: Scope as a Shield for Paradigm Protection

https://yueliusd.substack.com/p/commentary-on-materials-todays-rejection

Rethinking "Balanced View" in Scientific Controversies: Why Fairness Is Not Equivalence Between Correct and Incorrect Theories

https://yueliusd.substack.com/p/rethinking-balanced-view-in-scientific

Quick Decisions, Conventional Outcomes: How Rapid Editorial Processes Marginalize Disruptive Innovation

https://yueliusd.substack.com/p/quick-decisions-conventional-outcomes

The Editorial Orthodoxy in Academic Publishing: How Journals Favor Mainstream Conformity over Paradigmatic Innovation

https://yueliusd.substack.com/p/the-editorial-orthodoxy-in-academic

Yue Liu. Non-Mainstream Scientific Viewpoints in Microwave Absorption Research: Peer Review, Academic Integrity, and Cargo Cult Science, Preprints.org, preprint, 2025, DOI:10.20944/preprints202507.0015.v2, Supplementary Materials

Exposing Fundamental Misconceptions in Peer Review: A Critical Analysis of Editorial and Reviewer Failures in Microwave Absorption Theory Evaluation

A Critical Rebuttal to Systemic Reviewer and Editorial Errors in Microwave Absorption Research: Exposing Authority Bias, Scientific Misunderstanding, and the Failure of Peer Review

Commentary on Journal Rejections: The Liu et al. Microwave Absorption Theory Case

Challenging the Desk-Rejection Dogma

Publication Outlets for Sharp Criticism of Academia: A Deep Analysis of Institutional Gatekeeping and Systemic Suppression

Liu, Yue, Self-Citation Versus External Citation in Academic Publishing: A Critical Analysis of Citation Reliability, Publication Biases, And Scientific Quality Assessment (August 14, 2025). Available at SSRN: https://ssrn.com/abstract=5392646 or http://dx.doi.org/10.2139/ssrn.5392646

**Funding**

No funds, grants, or other support were received.

**Data Transparency**

Data sharing is not applicable as no datasets were generated or analyzed during the current study.

**Declarations**

**Competing Interests**

There are no relevant financial and non-financial interests to disclose.

**Ethics Approval for Research Involving Humans or Animals**

No human participants or animals were involved in this research

**References**

(1) Liu, Y.; Liu, Y.; Drew, M. G. B. Wave mechanics of microwave absorption in films - Distinguishing film from material. *Journal of Magnetism and Magnetic Materials* **2024**, *593*, 171850. DOI: 10.1016/j.jmmm.2024.171850.
(2) Liu, Y.; Liu, Y.; Drew, M. G. B. Wave mechanics of microwave absorption in films: A short review. *Optics and Laser Technology* **2024**, *178*, 111211. DOI: 10.1016/j.optlastec.2024.111211.
(3) Liu, Y.; Liu, Y.; Drew, M. G. B. Wave Mechanics of Microwave Absorption in Films: Multilayered Films. *Journal of Electronic Materials* **2024**, *53*, 8154–8170. DOI: 10.1007/s11664-024-11370-9.
(4) Liu, Y.; Yu, H.; Drew, M. G. B.; Liu, Y. A systemized parameter set applicable to microwave absorption for ferrite based materials. *Journal of Materials Science: Materials in Electronics* **2018**, *29* (2), 1562-1575. DOI: 10.1007/s10854-017-8066-0.
(5) Liu, Y.; Liu, Y.; Drew, M. G. B. A Re-evaluation of the mechanism of microwave absorption in film – Part 2: The real mechanism. *Materials Chemistry and Physics* **2022**, *291*, 126601. DOI: 10.1016/j.matchemphys.2022.126601.
(6) Liu, Y.; Ding, Y.; Liu, Y.; Drew, M. G. B. Unexpected Results in Microwave Absorption -- Part 1: Different absorption mechanisms for metal-backed film and for material. *Surfaces and Interfaces* **2023**, *40*, 103022. DOI: 10.1016/j.surfin.2023.103022.
(7) Liu, Y.; Drew, M. G. B.; Liu, Y. A physics investigation of impedance matching theory in microwave absorption film— Part 2: Problem analyses. *Journal of Applied Physics* **2023**, *134* (4), 045304. DOI: 10.1063/5.0153612.
(8) Liu, Y.; Liu, Y.; Drew, M. G. B. A theoretical investigation of the quarter-wavelength model-part 2: verification and extension. *Physica Scripta* **2022**, *97* (1), 015806. DOI: 10.1088/1402-4896/ac1eb1.
(9) Liu, Y.; Drew, M. G. B.; Liu, Y. A theoretical exploration of impedance matching coefficients for interfaces and films. *Applied Physics A* **2024**, *130*, 212. DOI:

10.1007/s00339-024-07364-3.
(10) Liu, Y.; Yin, X.; Drew, M. G. B.; Liu, Y. Reflection Loss is a Parameter for Film, not Material. *Non-Metallic Material Science* **2023**, *5* (1), 38 - 48. DOI: 10.30564/nmms.v5i1.5602.
(11) Liu, Y.; Drew, M. G. B.; Li, H.; Liu, Y. An experimental and theoretical investigation into methods concerned with "reflection loss" for microwave absorbing materials. *Materials Chemistry and Physics* **2020**, *243*, 122624. DOI: 10.1016/j.matchemphys.2020.122624.
(12) Liu, Y.; Lin, Y.; Zhao, K.; Drew, M. G. B.; Liu, Y. Microwave absorption properties of Ag/NiFe2-xCexO4 characterized by an alternative procedure rather than the main stream method using "reflection loss". *Materials Chemistry and Physics* **2020**, *243*, e122615. DOI: 10.1016/j.matchemphys.2019.122615.
(13) Liu, Y.; Liu, Y.; Drew, M. G. B. A re-evaluation of the mechanism of microwave absorption in film – Part 3: Inverse relationship. *Materials Chemistry and Physics* **2022**, *290*, 126521. DOI: 10.1016/j.matchemphys.2022.126521.
(14) Liu, Y.; Liu, Y.; Drew, M. G. B. A theoretical investigation on the quarter-wavelength model — Part 1： Analysis. *Physica Scripta* **2021**, *96* (12), 125003. DOI: <u>10.1088/1402-4896/ac1eb0</u>.
(15) Liu, Y.; Liu, Y.; Drew, M. G. B. Recognizing Problems in Publications Concerned with Microwave Absorption Film and Providing Corrections: A Focused Review. *Industrial & Engineering Chemistry Research* **2025**, *64* (7), 3635–3650. DOI: 10.1021/acs.iecr.4c04544.
(16) Liu, Y.; Ding, Y.; Liu, Y.; Drew, M. G. B. Unexpected results in Microwave absorption -- Part 2： Angular effects and the wave cancellation theory. *Surfaces and Interfaces* **2023**, *40*, 103024. DOI: 10.1016/j.surfin.2023.103024.
(17) Liu, Y.; Liu, Y.; Drew, M. G. B. Review of Wave Mechanics Theory for Microwave Absorption by Film. *Journal of Molecular Science* **2024**, *40* (4), 300-305. DOI: 10.13563/j.cnki.jmolsci.2023.06.018.
(18) Liu, Y.; Drew, M. G. B.; Liu, Y. A physics investigation of impedance matching theory in microwave absorption film—Part 1: Theory. *Journal of Applied Physics* **2023**, *134* (4), 045303. DOI: 10.1063/5.0153608.
(19) Liu, Y.; Drew, M. G. B.; Liu, Y. Theoretical insights manifested by wave mechanics theory of microwave absorption—Part 1: A theoretical perspective. *Preprints.org* **2025**. DOI: 10.20944/preprints202503.0314.v4.
(20) Liu, Y.; Liu, Y.; Drew, M. G. B. A Re-evaluation of the mechanism of microwave absorption in film – Part 1: Energy conservation. *Materials Chemistry and Physics* **2022**, *290*, 126576. DOI: 10.1016/j.matchemphys.2022.126576.
(21) Liu, Y.; Drew, M. G. B.; Liu, Y. Chapter 4: Fundamental Theory of Microwave Absorption for Films of Porous Nanocomposites: Role of Interfaces in Composite-Fillers. In *Porous Nanocomposites for Electromagnetic Interference Shielding*, Thomas, S., Paoloni, C., Pai, A. R. Eds.; Elsevier, 2024; pp 59 - 90.
(22) Liu, Y.; Drew, M. G. B.; Li, H.; Liu, Y. A theoretical analysis of the relationships shown from the general experimental results of scattering parameters s11 and s21 -- Exemplified by the film of BaFe12-iCeiO19/polypyrene with i = 0.2, 0.4, 0.6. *Journal of

*Microwave Power and Electromagnetic Energy* **2021**, *55* (3), 197–218. DOI: 10.1080/08327823.2021.1952835.

(23) Zhang, S.; Wang, T.; Gao, M.; Wang, P.; Pang, H.; Qiao, L.; Li, F. Strict proof and applicable range of the quarter-wavelength model for microwave absorbers. *Journal of Physics D: Applied Physics* **2020**, *53* (26), 265004. DOI: 10.1088/1361-6463/ab79da.

(24) Wang, X.; Du, Z.; Hou, M.; Ding, Z.; Jiang, C.; Huang, X.; Yue, J. Approximate solution of impedance matching for nonmagnetic homogeneous absorbing materials. *The European Physical Journal Special Topics* **2022**, *231* (24), 4213-4220. DOI: 10.1140/epjs/s11734-022-00570-1.

(25) Hou, Z.-L.; Gao, X.; Zhang, J.; Wang, G. A perspective on impedance matching and resonance absorption mechanism for electromagnetic wave absorbing. *Carbon* **2024**, *222*, 118935. DOI: 10.1016/j.carbon.2024.118935.

(26) Liu, Y. 推翻现行微波吸收理论的波动力学新理论的文章的他引情况*(Citation Situation of the Wave-mechanics New Theory Overturning Current Microwave Absorption Theory). ScienceNet.cn blog, 12 July 2025.* https://blog.sciencenet.cn/blog-3589443-1493361.html (accessed.

(27) Choi, J. R.; Cheon, S. J.; Lee, H. J.; Lee, S.-b.; Park, B.; Lee, H. Ultra-wideband electromagnetic wave absorption in mmWave using dual-loss engineered M-type hexaferrite: A wave cancellation approach. *Materials & Design* **2025**, *255*, 114220. DOI: 10.1016/j.matdes.2025.114220.

(28) Li, S.; Guo, S.; Chen, G.; Cui, Y.; Yang, H.; Qiu, J.; Wang, Z.; Dai, M.; Liu, S. Evolution of electron localization with Co2+ variations in Co Fe3−O4 hollow spheres for enhanced wave absorption. *Journal of Alloys and Compounds* **2025**, *1026*, 180381. DOI: 10.1016/j.jallcom.2025.180381.

(29) Zhou, Y.; He, P.; Ma, W.; Zuo, P.; Xu, J.; Tang, C.; Zhuang, Q. The Developed Wave Cancellation Theory Contributing to Understand Wave Absorption Mechanism of ZIF Derivatives with Controllable Electromagnetic Parameters. *Small* **2023**, 2305277. DOI: 10.1002/smll.202305277.

(30) Ray, S.; Panwar, R. Advances in Polymer-Based Microwave Absorbers—From Design Principles to Technological Breakthroughs: A Review. *IEEE Journal on Flexible Electronics* **2024**, *3* (9), 401-417. DOI: 10.1109/jflex.2024.3432103.

(31) Lu, J.; Xu, L.; Xie, C.; Zhang, C.; Han, Z.; Ren, Y.; Che, R. Microwave-Driven Dielectric-Magnetic Regulation of Graphite@alpha-MnO(2) Toward Enhanced Electromagnetic Wave Absorption. *Adv Sci (Weinh)* **2025**, e04489. DOI: 10.1002/advs.202504489    From NLM Publisher.

(32) Busti, N. D.; Parra, R.; Sousa Góes, M. Synthesis, Properties, and Applications of Iron Oxides: Versatility and Challenges. In *Functional Properties of Advanced Engineering Materials and Biomolecules*, Engineering Materials, 2021; pp 349-385.

(33) Abu Sanad, A. A.; Mahmud, M. N.; Ain, M. F.; Ahmad, M. A. B.; Yahaya, N. Z. B.; Mohamad Ariff, Z. Theory, Modeling, Measurement, and Testing of Electromagnetic Absorbers: A Review. *physica status solidi (a)* **2024**, *221* (4), 2300828. DOI: 10.1002/pssa.202300828.

(34) Xia, L.; Feng, Y.; Zhao, B. Intrinsic mechanism and multiphysics analysis of electromagnetic wave absorbing materials: New horizons and breakthrough. *Journal of*

*Materials Science & Technology* **2022**, *130*, 136-156. DOI: 10.1016/j.jmst.2022.05.010.
(35) Li, Q.; Zeng, Y.; Zhang, M.; Jiang, Z.; Xie, Z. The new insight into the microscopic enhancement mechanism of microwave absorption based on the electromagnetic heterogeneous interface of carbon nanocavity. *Journal of Colloid and Interface Science* **2025**, *699*, 138210. DOI: 10.1016/j.jcis.2025.138210.
(36) Green, M.; Chen, X. Recent progress of nanomaterials for microwave absorption. *Journal of Materiomics* **2019**, *5* (4), 503-541. DOI: 10.1016/j.jmat.2019.07.003.
(37) Liu, Y. Non-Mainstream Scientific Viewpoints in Microwave Absorption Research: Peer Review, Academic Integrity, and Cargo Cult Science. *preprints.org* **2025**. DOI: 10.20944/preprints202507.0015.v2
(38) Liu, Y. Scientific Accountability: The Case for Personal Responsibility in Academic Error Correction. *Qeios* **2025**, Preprint. DOI: 10.32388/m4ggkz.
(39) Liu, Y. Why Are Research Findings Supported by Experimental Data with High Probability Often False? --Critical Analysis of the Replication Crisis and Statistical Bias in Scientific Literature. *Preprints.org* **2025**. DOI: 10.20944/preprints202507.1953.v1.
(40) Liu, Y. The Entrenched Problems of Scientific Progress: An Analysis of Institutional Resistance and Systemic Barriers to Innovation. *Preprints.org* **2025**. DOI: 10.20944/preprints202507.2152.v1.
(41) Liu, Y.; Drew, M. G. B.; Liu, Y. Theoretical insights manifested by wave mechanics theory of microwave absorption — Part 2: A perspective based on the responses from DeepSeek. *Preprints.org* **2025**. DOI: 10.20944/preprints202503.0314.v3
(42) *Planck's principle*. https://en.wikipedia.org/wiki/Planck%27s_principle (accessed.
(43) Dave，X. *Why You Should Never Say "It's Just A Theory"*. 2016. https://www.youtube.com/watch?v=h0H-amOti_o (accessed 2025 18 January).
(44) Ziliak, S. T.; McCloskey, D. N. *The cult of statistical significance: how the standard error costs us jobs, justice, and lives*; The University of Michigan Press, 2008.
(45) *Non-mainstream viewpoints*. Liu, Yue. https://www.peeref.com/hubs/345 (accessed.
(46) Feynman, R. P.; Leighton, R. *"Surely You're Joking, Mr. Feynman", Adventures of a Curious Character*; W. W. Norton & Company, 2010.
(47) Weinstein, E. *The Problem With Peer Review*. YouTube, 2020. https://www.youtube.com/watch?v=U5sRYsMjiAQ (accessed 2025 1, Apr).